\documentclass{article}

\PassOptionsToPackage{numbers,sort&compress}{natbib}
\usepackage[final]{neurips_2023}
\usepackage[utf8]{inputenc} 
\usepackage[T1]{fontenc}    
\usepackage{hyperref}       
\usepackage{url}            
\usepackage{booktabs}       
\usepackage{amsfonts}       
\usepackage{nicefrac}       
\usepackage{microtype}      
\usepackage{xcolor}         

\usepackage{physics}
\usepackage{amsmath}

\usepackage{float}
\usepackage{caption}
\usepackage{subcaption}
\usepackage{graphicx}

\usepackage{tablefootnote}

\usepackage{siunitx} 
\usepackage{booktabs} 

\title{Accelerating Inference in Molecular Diffusion Models with Latent Representations of Protein Structure}

\author{%
  Ian Dunn \\
  Dept. of Computational \& Systems Biology\\
  University of Pittsburgh\\
  Pittsburgh, PA 15260 \\
  \texttt{ian.dunn@pitt.edu} \\
  \And
  David Ryan Koes \\
  Dept. of Computational \& Systems Biology\\
  University of Pittsburgh\\
  Pittsburgh, PA 15260 \\
  \texttt{dkoes@pitt.edu} \\
}

\begin{document}

\maketitle

\begin{abstract}
Diffusion generative models have emerged as a powerful framework for addressing problems in structural biology and structure-based drug design. These models operate directly on 3D molecular structures. Due to the unfavorable scaling of graph neural networks (GNNs) with graph size as well as the relatively slow inference speeds inherent to diffusion models, many existing molecular diffusion models rely on coarse-grained representations of protein structure to make training and inference feasible. However, such coarse-grained representations discard essential information for modeling molecular interactions and impair the quality of generated structures. In this work, we present a novel GNN-based architecture for learning latent representations of molecular structure. When trained end-to-end with a diffusion model for \textit{de novo} ligand design, our model achieves comparable performance to one with an all-atom protein representation while exhibiting a 3-fold reduction in inference time. \footnote{Our code is publicly available \url{https://github.com/dunni3/keypoint-diffusion}}
\end{abstract}

\section{Introduction}

There has been a surge of interest in leveraging diffusion models to address problems in structure-based drug design. These efforts have yielded promising outcomes, exemplified by successes in \textit{de novo} ligand design \cite{schneuing_structure-based_2023}, molecular docking \cite{corso_diffdock_2023}, fragment linker design \cite{igashov_equivariant_2022}, and scaffold hopping \cite{torge_diffhopp_2023}. These models apply diffusion processes on point cloud representations of protein/ligand complexes and employ geometric Graph Neural Networks (GNNs) to make denoising predictions. However, GNN memory and compute requirements scale unfavorably with graph size, and this scaling issue poses a particular challenge within diffusion models, due to their reliance on multiple forward passes to generate a single sample. Some of these molecular diffusion models use coarse-grained representations of their molecular systems to make training and inference computationally feasible \cite{corso_diffdock_2023,torge_diffhopp_2023}. While \cite{igashov_equivariant_2022} and \cite{schneuing_structure-based_2023} train models with both coarse-grained and all-atom protein representations, their results show superior performance when using all-atom representations at the cost of more expensive/time-consuming training and inference. This is likely because residue-level representations discard precise information regarding the orientation of side chains; information which is critical for modeling binding events \cite{he_alphafold2_2023,scardino_how_2023,masha_how_2023}.

In developing molecular diffusion models and applying them at scale, researchers must grapple with the trade off between the computational demands and performance afforded by their choice of molecular representation. This work proposes a new choice of molecular representation which simultaneously enjoys the expressiveness of all-atom representations and computational efficiency of coarse-grained representations. In summary, our main contributions are:


\begin{enumerate}
    \item A novel GNN-based architecture for learning condensed representations of molecular structure, allowing end-to-end training for downstream tasks that operate on these latent geometric representations.
    \item A diffusion model for \textit{de novo} ligand design that achieves a 3-fold increase in inference speed by conditioning ligand generation on a learned representation of protein structure.
\end{enumerate}



\section{Background}

\paragraph{Denoising diffusion probabilistic models}


Diffusion models \cite{ho_denoising_2020,kingma_variational_2022} define a forward diffusion process consisting of $T$ noising steps that convert samples from a data distribution at step $t=0$ to samples from a prior distribution at step $t=T$ by repeated additions of random noise. The forward diffusion process conditioned on an initial data point $x_0$ can be defined by Equation \ref{eq: foward_diff}. 

\begin{equation} \label{eq: foward_diff}
    q(x_t|x_0) = \mathcal{N}(x_t|\alpha_tx_0, \sigma_t^2\mathbb{I})
\end{equation}

Where $\alpha_t, \sigma_t \in \mathbb{R}^+$ are functions that control the amount of signal retained from and noise added to $x_0$, respectively. In this work, $\alpha_t$ is a function that smoothly transitions from $\alpha_0 \approx 1$ to $\alpha_T \approx 0$. We specifically work with variance-preserving diffusion processes for which $\alpha_t = \sqrt{1- \sigma_t^2}$. Equation \ref{eq: foward_diff} can be equivalently written as:

\begin{equation}
    x_t = \alpha_tx_0 +\sigma_t\epsilon \quad \epsilon \sim \mathcal{N}(0, \mathbb{I})
\end{equation}

A neural network that is trained to predict $\epsilon$ from noisy data points $x_t$ can be used to parameterize a reverse diffusion process $p_\theta(x_{t-1}|x_t)$ that converts samples from the prior distribution to samples from the training data distribution. We refer to this neural network as the ``noise prediction network'' $\hat{\epsilon}_\theta(x_t, t)$.

\paragraph{Equivariant diffusion on molecules}

\citet{hoogeboom_equivariant_2022} propose Equivariant Diffusion Models (EDMs) for generating 3D molecules. In this setting, a molecule with $N$ atoms is considered as a point cloud with positions $x \in \mathbb{R}^{N \times 3}$ and features $s \in \mathbb{R}^{N \times f}$, which are one-hot encoded atom types. For notational convenience, point clouds are represented with a single variable $z = [x, s]$. A forward diffusion process similar to that of Equation \ref{eq: foward_diff} is defined for both atom positions and features. The noise prediction network $\hat{\epsilon}_\theta(z_t, t)$ outputs an E(3)-equivariant vector $\hat{\epsilon}^{(x)}$ and E(3)-invariant vector $\hat{\epsilon}^{(s)}$ for each node, representing the noise to be removed from atom positions and features, respectively. 



\paragraph{Diffusion for protein-ligand complexes}

\citet{schneuing_structure-based_2023} introduce a conditional EDM for generating small molecules inside of a protein binding pocket, DiffSBDD. Both the ligand and the protein binding pocket are represented as point clouds $z^{(L)}$ and $z^{(P)}$, respectively. $z^{(L)}$ is an all-atom point cloud having one node per atom while $z^{(P)}$ is either an all-atom point cloud or a $C_\alpha$ point cloud containing one node for every residue in the binding pocket located at the alpha carbon position.

\citet{schneuing_structure-based_2023} propose two distinct diffusion processes for pocket-conditioned generation. The first is a conditional diffusion model where the diffusion process is defined only for $z^{(L)}$; the noise prediction network takes as input the noisy ligand $z_t^{(L)}$ and the pocket structure $z^{(P)}$ remains unchanged throughout the denoising process: $\hat{\epsilon}_\theta(z_t^{(L)}, z^{(P)}, t)$. The second method defines a joint diffusion process on both $z^{(L)}$ and $z^{(P)}$. $\hat{\epsilon}_\theta$ is trained to denoise both the ligand and pocket at every timestep $\hat{\epsilon}_\theta(z_t^{(L)}, z_t^{(P)}, t)$, and an inpainting procedure is used to generate ligands inside a given pocket. In both cases, $z^{(L)}$ and $z^{(P)}$ are passed to $\hat{\epsilon}_\theta$ as a heterogeneous graph where nodes are atoms or residues and edges are created based on euclidean distance between nodes.

\begin{figure}[pt]
     \centering
     \includegraphics[scale=1.0]{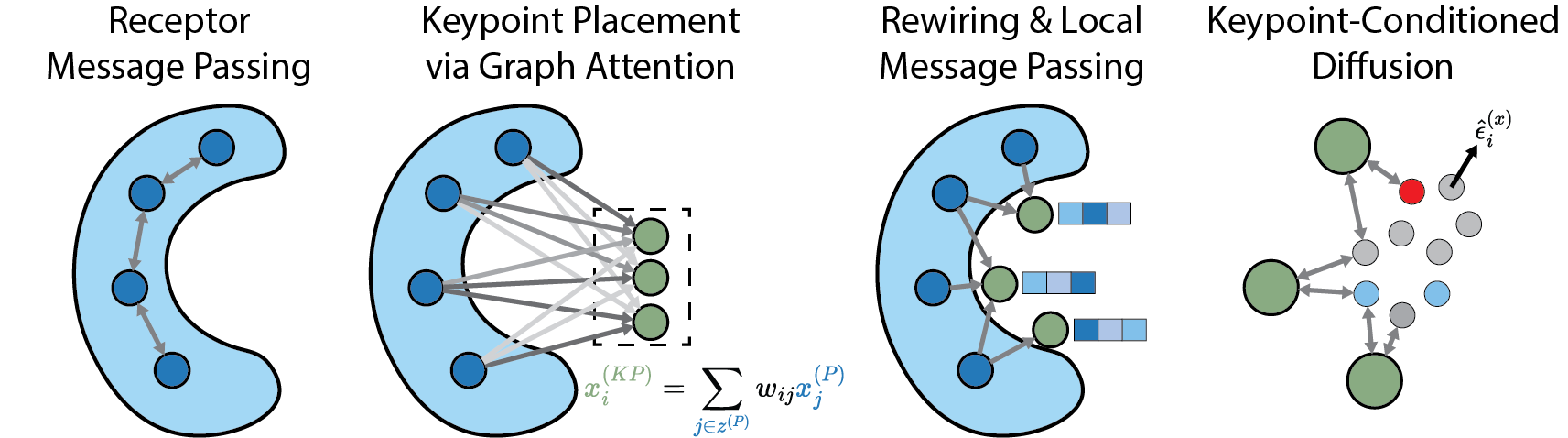}
     \caption{Message passing is performed between receptor nodes. Learned receptor embeddings are used to place keypoints inside the binding pocket. Keypoints extract local features of the binding pocket. Keypoints are then used to condition the ligand generation process.}
     \label{graphical_abstract}
\end{figure}

\section{Method}

To train a conditional EDM for pocket-conditioned ligand generation as described in \cite{schneuing_structure-based_2023}, the noise prediction network $\hat{\epsilon}_\theta$ must have access to some representation of the protein binding pocket. Taking inspiration from \citet{ganea_independent_2022}, we propose to use an encoder $E_\theta(z^{(P)})$ that accepts an all-atom protein point cloud as input and returns a small, fixed-size point cloud $z^{(KP)}$, which we term the ``keypoint representation''. The receptor encoder and diffusion model can be trained end-to-end by minimizing the denoising loss function Equation \ref{denoise_loss}.

\begin{equation} \label{denoise_loss} 
    \mathcal{L}_{DSM} =  \mathbb{E}_{t\sim U(0,T), z^{(L)}_0, z^{(P)}} \left[ \norm{\epsilon - \hat{\epsilon}_\theta \left(z_t^{(L)}, E_\theta(z^{(P)}), t \right) }^2 \right]
\end{equation}

DiffSBDD \cite{schneuing_structure-based_2023} and EDMs \cite{hoogeboom_equivariant_2022} parameterize $\hat{\epsilon}_\theta$ using the geometric GNN architecture known as EGNN \cite{satorras_en_2022}. Within the EGNN architecture, nodes possess a single vector feature that, in practice, is designated as the node's position in space. As a result, there is no point in the EGNN architecture where a node retains geometric information describing its local environment. We intuit that EGNN-based architectures may exhibit poor performance on structure representations where a node cannot be adequately described by a single point-mass i.e., residue or fragment point clouds. Specifically, we hypothesize that EGNN may struggle to learn representations of protein structure that are both informative and condensed. To investigate this phenomenon, we train all models with both EGNN and Geometric Vector Perceptron (GVP) \cite{jing_learning_2021,jing_equivariant_2021} based architectures. GVP-GNN can be seen as a generalization of EGNN to the setting where nodes can have an arbitrary number of vector features \cite{joshi_expressive_2023}.



\paragraph{Pocket encoder module}

The pocket encoder module is designed to take an all-atom point cloud of the protein binding pocket $z^{(P)}$ as input and produce a point cloud $z^{(KP)} = E_\theta(z^{(P)})$ having $K$ nodes as output. $K$ is a hyperparameter of the model chosen to be significantly smaller than the number of atoms in a binding pocket. In our training dataset, binding pockets have on the order of hundreds of nodes. We present results for models with $K = 40$ which is close to the average number of residues in a binding pocket. The nodes of $z^{(KP)}$, referred to as keypoints, have positions in space $x_i \in \mathbb{R}^3$ as well as scalar features $s_i \in \mathbb{R}^d$. When the pocket encoder module is parameterized with GVP-GNN, each keypoint is also endowed with vector features $v_i \in \mathbb{R}^{c \times 3}$. 

The sequence of operations within the pocket encoder module are summarized in Figure \ref{graphical_abstract}. First, message passing is performed along edges between binding pocket atoms. Keypoint nodes are then added to the graph without positions or features. Edges are drawn from receptor nodes to keypoint nodes to form a unidirectional complete bipartite graph. Keypoint positions are obtained via a dot-product variant of graph attention \cite{velickovic_graph_2018} along pocket-keypoint edges. Following keypoint position assignment is a ``graph rewiring'' step that selectively removes the aforementioned pocket-keypoint edges such that keypoints only have incoming edges from the \textit{nearest} pocket atoms. Finally, message passing along these local pocket-keypoint edges endows keypoint nodes with spatially localized features. Additional architectural details including equations for graph convolutions and keypoint placement are provided in Appendix \ref{appendix:architecture}.


\paragraph{Optimal transport loss}



We find that enforcing spatial alignment between keypoint positions and the true protein/ligand interface is a useful inductive bias. For each protein/ligand pair in the training set we compute a set of interface points $x^{(IP)} \in \mathbb{R}^{S \times 3}$ that are defined as the median points between all pairs of ligand atoms and binding pocket atoms $<5${\AA} apart. We apply an optimal transport loss function that is minimized when keypoint positions align with the true protein/ligand interface. 

\begin{equation} \label{myot}
    \mathcal{L}_{OT} = \underset{T \in \mathcal{U}(S,K)}{\min} \langle T, C \rangle \quad \textrm{where} \quad C_{s,k} = \norm{x^{(KP)}_k - x^{(IP)}_s}^2
\end{equation}

Where $\mathcal{U}(S,K)$ is the set of transport plans with uniform marginals and 
$\langle T, C \rangle$ is the Frobenius inner product between the transport map $T$ and the cost-matrix $C$. The optimal transport plan is solved in the forward pass using the python optimal transport package \cite{flamary_pot_2021} and is held fixed during the backwards pass.


\section{Results}

\paragraph{Experiments} We train all models on the BindingMOAD dataset \cite{hu_binding_2005} which contains approximately 40,000 experimentally determined protein/ligand structures from the Protein Data Bank \cite{berman_announcing_2003}. We train baseline models where the ligand point cloud is connected to the input protein point cloud without the use of any keypoint representation. Baseline models are trained with all-atom and $C_\alpha$ protein representations. We also train keypoint, all-atom, and $C_\alpha$ models with both EGNN and GVP architectures to evaluate the effect of GNN expressivity. 

We sample 100 ligands from every pocket in the test set. Generated ligands are subjected to a force-field minimization while holding the binding pocket fixed. We measure the RMSD of the ligand pose before and after minimization. If the ligand is in an unreasonable pose or forming unfavorable interactions with the binding pocket, there will be a larger RMSD upon minimization. Additionally, we use the Autodock Vina scoring function \cite{eberhardt_autodock_2021} to score the force-field minimized ligands and use the distribution of scores as a proxy for how well ligands are designed for their target pocket.

\begin{figure}[tbp]
     \centering
     \makebox[\textwidth][c]{\includegraphics[scale=0.99]{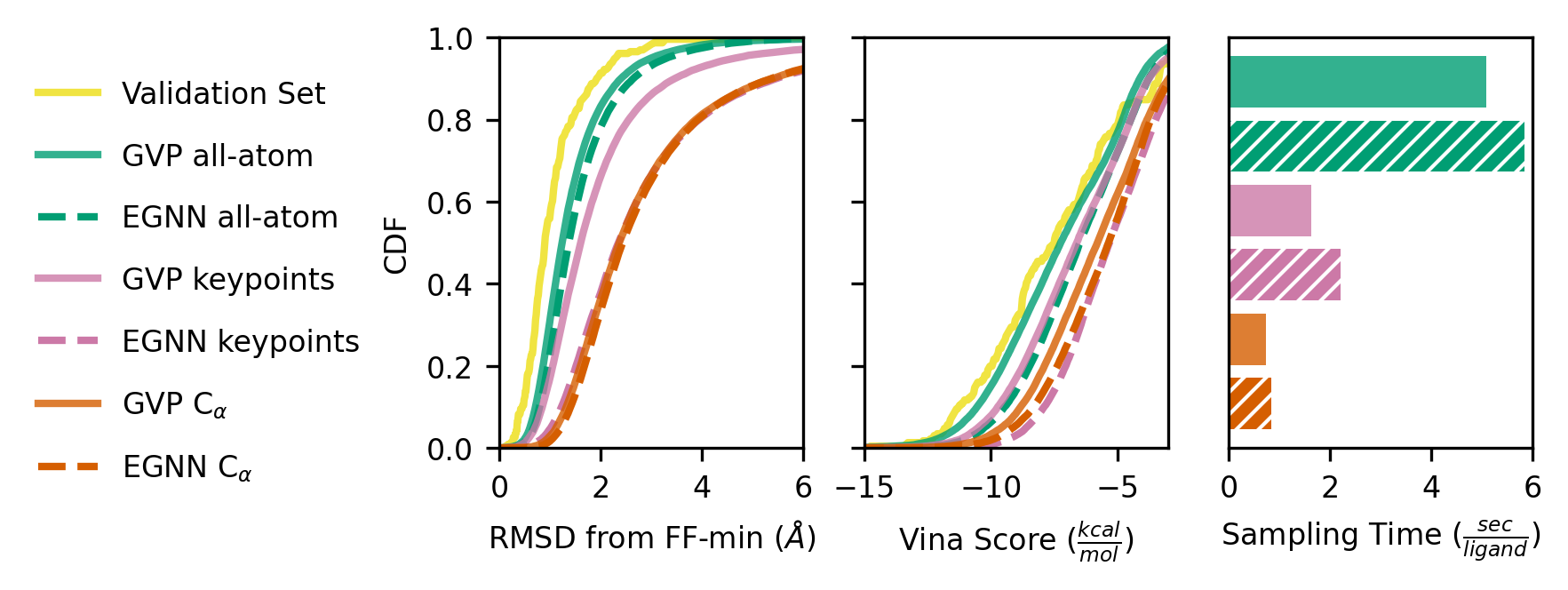}}
     \caption{Left, Middle: CDFs of ligand RMSD from force-field minimization and Vina score. Right: Sampling time per molecule averaged over the same ten binding pockets for each model.}
     \label{graphs}
\end{figure}

\paragraph{Generated Ligand Quality} We evaluate ligand quality by cumulative density functions of the RMSD from force-field minimization and Vina score shown in Figure \ref{graphs}, with higher CDF values indicating higher quality ligands for both metrics. Most notably, the GVP keypoint model performance is comparable to the all-atom models despite using ~10x fewer nodes to represent the binding pocket. Models using $C_\alpha$ binding pocket representations produce ligands of lower quality than those that use all-atom pocket representations; this is consistent with prior works \cite{schneuing_structure-based_2023,igashov_equivariant_2022}. The EGNN keypoint model produces ligands of equivalent quality to that of $C_\alpha$ models. 



\paragraph{Inference Performance}

We sample 100 molecules per pocket for 10 binding pockets and report mean wall-time per binding pocket. Sampling times in Figure \ref{graphs} show that keypoint models are ~3x faster than their corresponding all-atom models. Additional results in Appendix \ref{appendix:varying_keypoints} show that we can trade-off inference time and ligand quality by changing the number of keypoints.

\section{Conclusions}

Our receptor encoder module is capable of learning compressed representations of binding pocket structure which enables a 3x reduction in inference time while maintaining comparable quality of generated ligands. Our receptor encoder module may serve as a useful tool for scaling inference in molecular diffusion models. Moreover, our work demonstrates that learned structure encoders can provide valuable flexibility to trade-off computational demands and model performance.

The GVP keypoint model was able to approach all-atom levels of performance while the EGNN keypoint model failed to exceed the performance $C_\alpha$ representations. This result supports our hypothesis that EGNN struggles to learn on molecular representations where a single node represents multiple atoms and may serve as practical guidance for practitioners designing geometric deep learning models for molecular structure. 

\subsection*{Funding}
This work is funded through R35GM140753 from the National Institute of General Medical Sciences. The content is solely the responsibility of the authors and does not necessarily represent the official views of the National Institute of General Medical Sciences or the National Institutes of Health.

\typeout{} 
\bibliography{refs}

\appendix







\section{Additional Experimental Details}

We use the BindingMOAD dataset splits provided by \citet{schneuing_structure-based_2023}.

Binding pockets are defined as the set of residues having any atoms within $8$\AA of any ligand atoms.

Models are trained and sampled with $T=1000$ diffusion steps and the same polynomial noise schedule used in \cite{hoogeboom_equivariant_2022,schneuing_structure-based_2023}.

Force-field minimization is done using the UFF force-field implemented in RDKit \cite{noauthor_rdkit_nodate}.

For inference time analysis, inference is run using the same batch size for all models on the same GPU. 

The baseline models are conceptually/functionally equivalent to DiffSBDD; however, we do not use their implementation. The baseline models are our re-implementation of DiffSBDD; this enables meaningful comparisons that control for factors outside of the protein representation e.g., minor architectural differences, inference time differences due to code efficiency. 

\section{Architecture} \label{appendix:architecture}

We operate on point clouds. Every point in the cloud is a node. Specifically, there are three distinct point clouds present in this work: the ligand point cloud $z^{(L)}$, the protein binding pocket point cloud $z^{(P)}$, and the keypoint point cloud produced by the receptor encoder $z^{(KP)} = E_\theta(z^{(P)})$. These three point clouds are combined into a single heterogeneous graph having three node types: protein nodes, keypoint nodes, and ligand nodes. Edges in the graph are directed, resulting in $9$ possible unique edge types. 

When using the EGNN architecture, each node $i$ in the point cloud is endowed with a position in space $x_i \in \mathbb{R}^3$ and scalar features $s_i \in \mathbb{R}^d$. For all-atom point clouds, $s_i$ are initialized as one-hot vectors of atom elements. For $C_\alpha$ point clouds, $s_i$ are initialized as one-hot vectors of the amino acid identity. When using the GVP-GNN architecture, nodes are additionally endowed with vector features $v_i \in \mathbb{R}^{c \times 3}$ which are initialized to zeros.

A graph convolution is defined as one round of message passing, message aggregation, and node feature updates. A single graph convolution makes up one layer of a GNN. Both the receptor encoder and the noise prediction models stack several graph convolution layers.

Separate message-generating and node-update functions are instantiated for each edge and node type, each having their own set of learnable parameters. Within the receptor-encoder, graph convolutions are only done along one edge type at a time: $P \to P$ edges followed by $KP \to P$ edges. Within the noise prediction network, graph convolutions are performed along the following edge types simultaneously: $KP \to L$, $L \to KP$, $KP \to KP$, and $L \to L$.

\subsection{EGNN graph convolution}

EGNN computes separate messages for scalar and position features, which are defined by Equations \ref{egnn_mij_scalar} and \ref{egnn_mij_position}, respectively. 

\begin{equation}
    \label{egnn_mij_scalar}
    m_{i \to j}^{(s)} = \phi_s(s_i, s_j, d_{ij}) 
\end{equation}

\begin{equation}
    \label{egnn_mij_position}
    m_{i \to j}^{(x)} = \frac{x_i - x_j}{d_{ij}}\phi_x(s_i, s_j,d_{ij})
\end{equation}

Where $d_{ij}$ is the Euclidean distance between nodes $i$ and $j$. The functions $\phi_s: \mathbb{R}^{2d+1} \to \mathbb{R}^{d}$ and $\phi_x: \mathbb{R}^{2d+1} \to \mathbb{R}$ are implemented as shallow multi-layer perceptrons (MLPs). Incoming messages are aggregated on each node and used to update the scalar features and positions of each node. 

\begin{equation}
    \label{egnn_update_scalar}
    s_i^{(l+1)} = s_i^{(l)} + \frac{1}{C}\sum_{j \in \mathcal{N}(i)} m_{j \to i}^{(s)}
\end{equation}

\begin{equation}
    \label{egnn_update_position}
    x_i^{(l+1)} = x_i^{(l)} + \frac{1}{C}\sum_{j \in \mathcal{N}(i)} m_{j \to i}^{(x)}
\end{equation}

Note that we are overloading the node feature notation here. Previously the superscript above a node feature was used to indicate the type of node (ligand, keypoint, protein). In Equations \ref{egnn_update_scalar} and \ref{egnn_update_position}, the superscript above node features indicates the layer of network: $s_i^{(l)}$ and $x_i^{(l)}$ are node $i$'s scalar and position features at layer $l$. $C$ is average in-degree of the graph, and $\mathcal{N}(i)$ is the set nodes which have edges pointing to node $i$.

Layer normalization is applied to scalar features after each graph convolution.

For a noise prediction network containing $L$ EGNN convolutions, the predicted position noise is obtained by subtracting the layer-$0$ node positions from the layer-$L$ node positions: $\hat{\epsilon}^{(x)} = x^{(L)} - x^{(0)}$. The predicted noise for atom features is obtained by passing the scalar features for each node at layer $L$ through a shallow MLP with a linear output layer:

\begin{equation}
    \hat{\epsilon}^{(s)} = \phi(s^{(L)})
\end{equation}

\subsection{GVP-GNN graph convolution}

Our implementation of GVP-based graph convolutions do not update node positions; rather, they update scalar and vector features. GVPs accept and return a tuple of scalar and vector features. Therefore, scalar and vector messages $m_{i \to j}^{(s)}$ and $m_{i \to j}^{(v)}$ are generated by a single function $g_m$ which is two GVPs chained together.

\begin{equation}
    \label{gvp_mij}
    m_{i \to j}^{(s)}, m_{i \to j}^{(v)} = g_m\left([s_i : rbf(d_{ij})], \left[v_i : \frac{x_i - x_j}{d_{ij}} \right]\right) 
\end{equation}

Where $:$ denotes concatenation, and $rbf(d_{ij})$ is a radial basis function (RBF) embedding of the distance between nodes $i$ and $j$.

The node-update function is defined exactly as described in \cite{jing_equivariant_2021}:

\begin{equation}
    \label{gvp_update}
    s_i^{(l+1)}, v_i^{(l+1)} = \mathrm{LN} \left( [s_i^{(l)}, v_i^{(l)}] + \\
    \mathrm{DO}\left( g_u \left( \frac{1}{|\mathcal{N}(i)|} \sum_{j \in \mathcal{N}(i) } \left[ m_{j \to i}^{(s)}, m_{j \to i}^{(v)} \right]   \right) \right) \right)
\end{equation}

Where LN and DO are dropout and layer norm, respectively. The node update function $g_u$ is a chain of three GVPs.

\subsection{GVP noise-prediction block}

After GVP graph convolutions, every ligand node is in possession of scalar and vector features, $s_i$ and $v_i$, which contain information about the ligand atom's environment; these features for each ligand atom are passed into ``noise-prediction block'' which returns noise predictions $\hat{\epsilon}^{(s)}$, $\hat{\epsilon}^{(x)}$.

Within the noise prediction block, $s_i$ and $v_i$ are first passed through a chain of 4 GVPs. The first three GVPs output the same number of scalar and vector features as are in the input. The final GVP outputs the input number of scalar features but only 1 vector feature and the vector-gating activation function, which is typically a sigmoid function, is replaced with the identity. The output vector feature is $\hat{\epsilon}^{(x)}$. The output scalar features are passed through an additional MLP to produce a scalar vector of the same shape as the one-hot encoded atom type vectors; this output becomes $\hat{\epsilon}^{(s)}$.

\subsection{Keypoint placement via graph attention}

After several graph convolutions along $P \to P$ edges, $z^{(P)}$ nodes possess features describing their local atomic environment which will be used to determine the positions of keypoints. $K$ keypoints are added to the graph but they initially have no positions, scalar features, or vector features associated with them. Each node in $z^{(P)}$ is a given an out-going edge to every keypoint; thus the nodes in $z^{(P)}$ and $z^{(KP)}$ form a unidirectional complete bipartite graph. 

Keypoint positions are then computed as a weighted sum of pocket node positions.

\begin{equation}
    x_i^{(KP)} = \sum_{j \in z^{(P)}} w_{ij}x_j^{(P)} 
\end{equation}

\begin{equation}
    w_{ij} = \frac{\tilde{w}_{ij}}{\sum_{k \in z^{(P)}} \tilde{w}_{ik} }
\end{equation}

\begin{equation}
    \tilde{w}_{ij} = \exp{ \phi_i(\mu(s^{(P)})) \cdot \phi_P(s_j^{(P)}) }
\end{equation}

Where $\mu$ takes the mean of all pocket atom scalar features. $\phi_i$ is a shallow MLP that is unique to each keypoint. $\phi_P$ is also a shallow MLP. This mechanism can be thought of as a dot-product graph attention mechanism where the ``keys'' and ``querys'' are some function of the source and destination node features and the ``values'' are the positions of the source node.

\section{Model Hyperparameters}

The receptor encoder module is implemented with 4 graph convolutions on $P \to P$ edges and two graph convolutions on $P \to KP$ edges following keypoint placement / graph rewiring. Denoising diffusion models contain 6 graph convolutions.

All nodes in all models are with 256 scalar features. GVP-based models use 16 vector features per node.

\begin{table}[h]
    \centering
    \caption{Edge Assignment Methods}
    \label{kp_edge_table}
    \begin{tabular}{l|ll}
         \toprule
         Edge Type &  Keypoint Models & All-Atom \& $C_\alpha$ \\
         \midrule
         $P \to P$   & $d_{ij} \leq 3.5$\AA & $d_{ij} \leq 3.5$\AA \\
         $P \to KP$  \tablefootnote{Pocket-keypoint edges created only after keypoint positions are determined.}
         & 5-NN & - \\
         $P \to L$ & - & 5-NN \\
         $KP \to KP$ & $d_{ij} \leq 8$\AA & - \\
         $KP \to L$ & 5-NN & - \\
         $L \to L$ & $d_{ij} \leq 5 $ \AA & $d_{ij} \leq 5 $ \AA  \\
    \end{tabular}
\end{table}

\section{Additional Results}

In the main body of the paper, we report keypoint models that have been trained with $K=40$ keypoints. In this section we also report results for keypoint models trained with $K=20$ keypoints. These models are denoted by the names ``40kp'' and ``20kp'', respectively. Models with more keypoints give slightly better quality ligands at the cost of longer inference times.

\subsection{Main results in table format}

\begin{tabular}{lrrr}
\toprule
                   Model &  \% w/ RMSD < 2 &  \% w/ Vina Score < -7 $\frac{kcal}{mol}$ &  Median Sampling Time $\left(\frac{s}{ligand}\right)$ \\
\midrule
          Validation Set &          91.34 &                 57.14 &                   - \\
            GVP all-atom &          83.33 &                 52.83 &                  5.09 \\
           EGNN all-atom &          78.28 &                 41.38 &                  5.85 \\
                GVP 20kp &          60.01 &                 38.78 &                  0.98 \\
                GVP 40kp &          66.36 &                 44.10 &                  1.63 \\
               EGNN 20kp &          29.86 &                 17.45 &                  1.07 \\
               EGNN 40kp &          37.95 &                 20.49 &                  2.21 \\
 GVP $C_\alpha$ &          36.51 &                 32.11 &                  0.74 \\
EGNN $C_\alpha$ &          33.56 &                 24.77 &                  0.84 \\
\bottomrule
\end{tabular}

\subsection{Figures for varying keypoint number} \label{appendix:varying_keypoints}

\begin{figure}[H]
     \centering
     \makebox[\textwidth][c]{\includegraphics[scale=0.99]{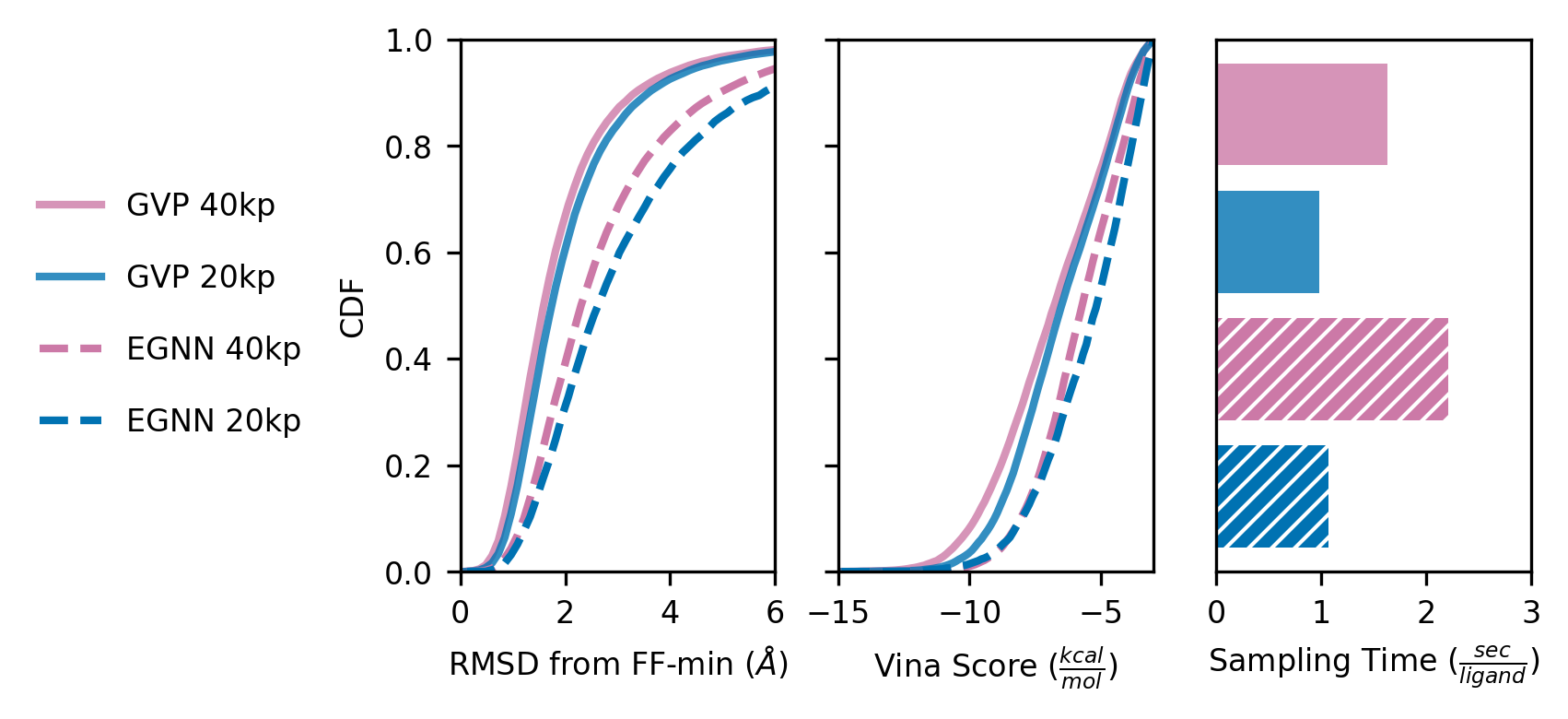}}
     \caption{Left, Middle: CDFs of ligand RMSD from force-field minimization and Vina score. Right: Sampling time per molecule averaged over the same ten binding pockets for each model.}
     \label{graphs_multikp}
\end{figure}

\subsection{Additional ligand quality metrics}

Prior works on \textit{de novo} ligand generation have reported QED (drug-likeness), SA (synthetic accessibility), and diversity metrics on generated ligands. Although we maintain that ligand-only metrics are less critical for evaluating generative models than metrics that describe ligand/pocket interactions, we nevertheless provide these metrics for ligands generated by our model in Table \ref{table:extra_metrics}. Diversity is the average value of the complement of the Tanimoto score over all pairs of ligands generated for a given pocket.

\begin{table}[H]
    \centering
    \caption{Additional Ligand Quality Metrics}
    \begin{tabular}{l|cccc}
        \hline
         & QED & SA & Diversity \\
                     & (Mean $\pm$ Std) & (Mean $\pm$ Std) & (Mean $\pm$ Std) \\
        \hline
        Validation Set & 0.59 $\pm$ 0.13 & 0.37 $\pm$ 0.09 & - \\
        EGNN all-atom & 0.57 $\pm$ 0.13 & 0.34 $\pm$ 0.09 & 0.77 $\pm$ 0.06 \\
        GVP all-atom & 0.61 $\pm$ 0.14 & 0.34 $\pm$ 0.09 & 0.73 $\pm$ 0.10 \\
        EGNN 20kp & 0.60 $\pm$ 0.13 & 0.34 $\pm$ 0.08 & 0.79 $\pm$ 0.05 \\
        EGNN 40kp & 0.56 $\pm$ 0.14 & 0.35 $\pm$ 0.08 & 0.78 $\pm$ 0.06 \\
        GVP 20kp & 0.61 $\pm$ 0.14 & 0.34 $\pm$ 0.08 & 0.74 $\pm$ 0.08 \\
        GVP 40kp & 0.61 $\pm$ 0.13 & 0.34 $\pm$ 0.08 & 0.73 $\pm$ 0.10 \\
        EGNN $C_\alpha$ & 0.59 $\pm$ 0.13 & 0.33 $\pm$ 0.09 & 0.78 $\pm$ 0.06 \\
        GVP $C_\alpha$ & 0.62 $\pm$ 0.14 & 0.34 $\pm$ 0.08 & 0.77 $\pm$ 0.07 \\
        \hline
    \end{tabular}
    \label{table:extra_metrics}
\end{table}

\end{document}